\documentclass[aps,prl,reprint]{revtex4-2}
\usepackage{amsmath}
\usepackage{amssymb}
\usepackage{graphicx}
\usepackage{hyperref}
\usepackage{url}
\usepackage{color,xcolor}
\DeclareMathOperator{\sech}{sech}
\DeclareMathOperator{\atan}{atan}

\begin{document}
\title{Manipulation of magnetic domain wall by ferroelectric switching: Dynamic magnetoelectricity at the nanoscale}
\author{Jun Chen}
\author{Shuai Dong}
\email{Corresponding author. Email: sdong@seu.edu.cn}
\affiliation{School of Physics, Southeast University, Nanjing 211189, China}
\date{\today}

\begin{abstract}
Controlling magnetism using voltage is highly desired for applications, but remains challenging due to fundamental contradiction between polarity and magnetism. Here we propose a mechanism to manipulate magnetic domain walls in ferrimagnetic or ferromagnetic multiferroics using electric field. Different from those studies based on static domain-level couplings, here the magnetoelectric coupling relies on the collaborative spin dynamics around domain walls. Accompanying the reversal of spin chirality driven by polarization switching, a ``rolling-downhill"-like motion of domain wall is achieved at the nanoscale, which tunes the magnetization locally. Our mechanism opens an alternative route to pursuit practical and fast converse magnetoelectric functions via spin dynamics.
\end{abstract}
\maketitle

A vital but challenging question for spintronics is how to effectively tune magnetism of materials using electric voltage, which can reduce the energy consumption, raise the storage density, speed up the data writing, and thus lead to a revolution of magnetism-based information storage and operation. Multiferroics, with coexisting magnetism and polarity in single phase materials, provide an ideal platform to pursuit such converse magnetoelectric (CME) effect. In the past decades, the frontiers of magnetoelectricity have been greatly pushed forward, with many new materials discovered, new mechanisms revealed, and new devices demonstrated \cite{dong_multiferroic_2015,
fiebig_evolution_2016,dong_magnetoelectricity_2019,spaldin_advances_2019,lu_single-phase_2019}.

One focused direction of CME is to use the ferroelectric domains to lock proximate ferromagnetic domains, as done in BiFeO$_3$- or BaTiO$_3$-based heterostructures \cite{chu_electric-field_2008,bea_mechanisms_2008,heron_deterministic_2014,franke_reversible_2015,you_uniaxial_2010}. Its mechanism mainly relies on the ferroelasticity and magnetocrystalline anisotropy \cite{heron_electric_2014}. Usually a ferroelectric domain is also a ferroelastic domain, thus the easy axis of proximate soft magnet can be modulated accompanying the ferroelastic switching, resulting in the rotating of magnetization within magnetic domain. In principle, such magnetoelectric coupling belongs to the canonical type, expressed as $(\textbf{M}\cdot\textbf{P})^2$ ($\textbf{M}$/$\textbf{P}$: magnetic/polar order parameters). This CME does not work for the $180^\circ$ ferroelectric switching (or $180^\circ$ ferroelectric domain wall), as a characteristic of ferroelasticity. Another direction to pursuit CME is based on the ferroelectric field effect \cite{rondinelli_carrier-mediated_2008,burton_prediction_2009,garcia_ferroelectric_2010,dong_full_2013}, which can be expressed as a coupling like $(\nabla\cdot\textbf{P})(\textbf{M})^2$ or its variants \cite{weng_inversion_2016}. Both these routes are static couplings in the domain level \cite{fusil2014magnetoelectric}.

In this Letter, a conceptually distinct mechanism is proposed to realize significant CME via the motion of chiral ferrimagnetic (or ferromagnetic) domain walls in multiferroics, which can be controlled by electrical field. The chirality of magnetic domain wall commonly exists in magnets \cite{,cheong2019sos}, but its vital role to CME was mostly omitted in past studies \cite{khomskii_classifying_2009}. Different from aforementioned CME relying on static couplings of magnetism and polarity within or between domains, here the spin dynamic process around domain wall is the key ingredient. In this sense, our mechanism mimics the magnetic racetrack memory \cite{parkin_magnetic_2008}, but its driving force is the Dzyaloshinskii-Moriya (DM) interaction, instead of the spin current.

\begin{figure}
\includegraphics[width=0.48\textwidth]{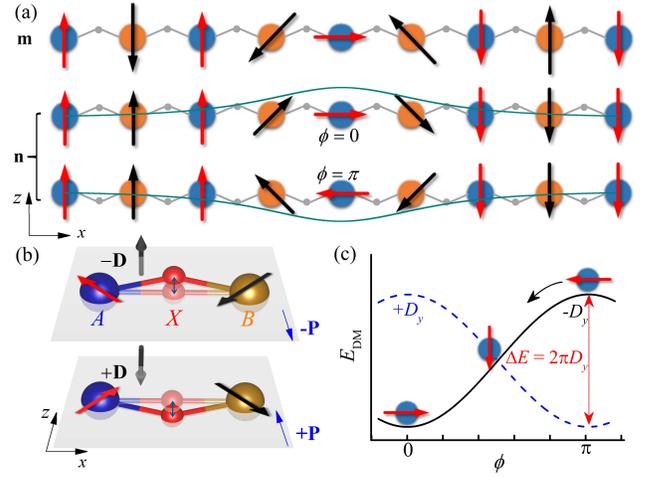}
\caption{(a) Schematic of ferrimagnetic domain walls. Magnetic sublattices $A$ and $B$ are distinguished by colors. $\textbf{m}$: magnetic moments. $\textbf{n}$: normalized staggered moments defined as $\frac{\textbf{m}^A}{|\textbf{m}^A|}$ and $-\frac{\textbf{m}^B}{|\textbf{m}^B|}$. The chirality of domain wall is characterized by the azimuthal angle of $\textbf{n}$ at the center of wall: clockwise ($\phi=0$) or counterclockwise ($\phi=\pi$). (b) The major component of DM vector $\textbf{D}$ is perpendicular to the $A$-$X$-$B$ plane, whose sign is determined by the displacement direction of anion $X$. (c) Schematic of ``rolling-downhill"-like mechanism. The domain wall energy from DM interaction is a function of chirality (i.e. $\phi$). For given chirality, the reversal of DM vector makes the energy minimum to be maximum, which leads to the spin dynamics of domain wall.}
\label{schematic}
\end{figure}

\textit{Model.} Let's start from the simplest case, an one-dimensional ferrimagnetic chain along the $x$-axis [Fig.~\ref{schematic}(a)]. The ferromagnetic and antiferromagnetic cases can be covered by extending the ferrimagnetism to two end limits. The magnetic moments at sublattices A and B are $\textbf{m}^A$ and $\textbf{m}^B$. All nearest-neighbor $\textbf{m}^A$ and $\textbf{m}^B$ are coupled by antiferromagnetic exchange as $J\textbf{s}^A\cdot\textbf{s}^B$, where $\textbf{s}$ is the normalized vector $\textbf{m}/|\textbf{m}|$ and $J$ is positive. The magnetic easy axis is assumed to along the $z$-axis with single-axis magnetocrystalline anisotropy (energy item: $-Ks_z^2$, where $K$ is positive and $s_z$ is the $z$-component of $\textbf{s}$).

If this system is multiferroic with polarization along the $z$-axis, the DM interaction in the form of $\textbf{D}\cdot(\textbf{s}^A\times\textbf{s}^B$) will be generated due to spin-orbit coupling (SOC). The vector $\textbf{D}$ is along the $y$-axis and its sign is determined by the direction of polarization [Fig.~\ref{schematic}(b)] \cite{moriya1960anisotropic,dong_magnetoelectricity_2019}. This DM interaction prefers a noncollinear cycloid order with moments rotating in the $xz$ plane.

Thus, the Hamiltonian of ferrimagnetic chain reads as:
\begin{eqnarray}
\nonumber H&=&\sum_i{J(\textbf{s}_i^A\cdot\textbf{s}_i^B+\textbf{s}_i^B\cdot\textbf{s}_{i+1}^A)}+
\textbf{D}\cdot(\textbf{s}_i^A\times\textbf{s}_i^B+\\
&&\textbf{s}_i^B\times\textbf{s}_{i+1}^A)-[K^A(s_{i,z}^A)^2+K^B(s_{i,z}^B)^2],
\label{hamid}
\end{eqnarray}
where $i$ is the index of two-site unit cell. If the magnetocrystalline anisotropy and exchange are strong enough (e.g. $8KJ>\pi^2|\textbf{D}|^2$ if $K^A$=$K^B$), the collinear texture will be stable \cite{bogdanov_magnetic_2002}. In the following, this collinear limit with the $K^A$=$K^B$ approximation will be considered. In the continuous limit, Eq.~\ref{hamid} can be rewritten as:
\begin{equation}
H=\int{[\frac{J}{2}(\nabla\textbf{n})^2+\textbf{D}\cdot(\textbf{n}\times\nabla\textbf{n})-Kn_z^2]}dx,
\label{hamic}
\end{equation}
where $\textbf{n}$ is the normalized staggered moments [Fig.~\ref{schematic}(a)].

Despite the collinearity within each domain, noncollinear magnetic textures exist at domain walls. Here an $180^\circ$ N\'eel-type domain wall (due to the DM term) is considered, whose chirality can be described by the azimuthal angle $\phi$ of $\textbf{n}$ at the center of wall [Fig.~\ref{schematic}(a)]. By minimizing the energy (Eq.~\ref{hamic}) in the small $\textbf{D}$ limit, the analytical solution for an isolated domain wall is  \cite{leeuw_dynamic_1980}:
\begin{equation}
\sin\theta=\sech(x/\Delta),
\label{theta}
\end{equation}
where $\Delta=\sqrt{J/2K}$, $\theta$ is the polar angle of $\textbf{n}$, and the center of wall is at $x=0$. The topological charge for such a domain wall can be defined as $
C=\frac{1}{\pi}\int_{-\infty}^{\infty}\nabla\theta dx$ which equals $1$ or $-1$, depending on the $\textbf{n}$ texture (from $+z$ to $-z$, or $-z$ to $+z$). The integral DM energy of domain wall in the continuous limit is:
\begin{equation}
E_{\rm DM}=\pi D_y\cos\phi.
\end{equation}
Thus the energy difference between clockwise and counterclockwise chiral domains are $2\pi D_y$ [Fig.~\ref{schematic}(c)]. Then, after the polarization (i.e., the sign of $D_y$) reversal, the chirality of domain wall should reverse via an energy dissipative process of spin dynamics.

\begin{figure}
\includegraphics[width=0.48\textwidth]{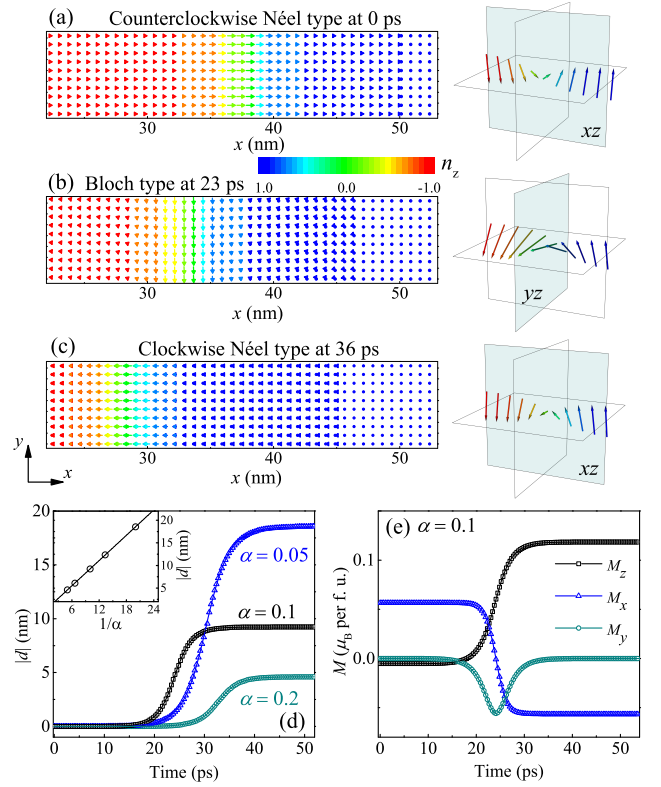}
\caption{(a-c) Snapshots of $\textbf{n}$ around the domain wall at $0$ ps (counterclockwise N\'eel-type), $23$ ps (intermediate Bloch-type), and $36$ ps (clockwise N\'eel-type), since the sudden reversal of $D_y$. Left: top view. The in-plane $xy$-components are represented by arrows and the $z$-components are indicated by colormap. Right: corresponding side view. (d) Motion distance ($|d|$) of domain wall center as a function of time with various $\alpha$'s. Insert: the relationship between final distance $|d_s|$ and $\frac{1}{\alpha}$. (e) The $x$-, $y$-, and $z$-components of net magnetization (in average of the whole lattice) as a function of time. The change of $z$-component is due to the resize of domains, while the reversal of $x$-component is due to the reversal of chirality. The emergence of $y$-component is due to the  intermediate Bloch-type state. In (d) and (e), the simulation results (dots) agree with the analytical solutions (curves) well.}
\label{md}
\end{figure}

\textit{Dynamics.} To investigate the domain wall dynamics of chirality reversing, the Landau-Lifshitz-Gilbert (LLG) equation is employed \cite{evans_atomistic_2014}, which reads as:
\begin{eqnarray}
\frac{\partial\textbf{m}}{\partial{t}}=-\gamma(\textbf{m}\times\textbf{f})+\alpha(\textbf{m}\times\frac{\partial\textbf{m}}{\partial{t}}),
\label{LLG}
\end{eqnarray}
where the effective field $\textbf{f}$=$-\partial{H}/\partial{\textbf{m}}$, the gyromagnetic ratio $\gamma$=$g\mu_{\rm B}/\hbar$, and $\alpha$ is the Gilbert damping coefficient.

In our following simulations, all coefficients except $\alpha$ are taken from the density functional theory (DFT) calculation of BiFe$_{1/2}$Co$_{1/2}$O$_3$ \cite{SM}, which is a typical ferroelectric ferrimagnet as required \cite{gao_room-temperature_2018,menendez_giant_2020}. Our physical mechanism and conclusion are not limited to a specific material but should be generally valid.

Using the fourth-order Runge-Kutta method, the LLG simulation is performed on a $200\times20$ rectangular lattice. The periodic boundary conditions are imposed in the $y$ direction, while the open boundary conditions are set in the $x$ direction and the moments at two ends are fixed. Starting from a sharp wall, an isolated domain wall is relaxed first, and the optimized texture matches Eq.~\ref{theta}.

The reversal of $\textbf{D}$ vector breaks the stability of chiral domain wall and leads to the gradual reversal of chirality via a dissipative process. During this process, the domain wall deforms from the N\'eel-type $\phi=0$ wall [Fig.~\ref{md}(a)] to the N\'eel-type $\phi=\pi$ wall [Fig.~\ref{md}(c)] via the intermediate Bloch-type wall  [Fig.~\ref{md}(b)].

Interestingly, a translational motion of domain wall occurs accompanying the chirality reversal, as shown in Fig.~\ref{md}(d). In the initial stage, the process starts slowly, then speeds up in the intermediate stage, and finally slows down to stop. The whole process is an analogy of ``rolling-downhill" with resistance [Fig.~\ref{schematic}(c)]. Using a relative large damping coefficient $\alpha$ (e.g. $0.1$), the movement of domain wall can reach $\sim9$ nm. Noting that the final distance $d_s$ is determined by the dampling: the weaker $\alpha$ the farther $d_s$ [insert of Fig.~\ref{md}(d)], which can be intuitionally understood as an energy dissipative motion.

The motion of domain wall resizes domains, and thus changes the $z$-component of magnetization [Fig.~\ref{md}(e)]. Also, the $x$-component of magnetization (coming from the domain wall) reverses following the chirality reversal. Meanwhile, the $y$-component emerges during the motion and fades away when the motion stops [Fig.~\ref{md}(e)], which is a characteristic of intermediate Bloch-type wall.

\begin{figure}
\includegraphics[width=0.48\textwidth]{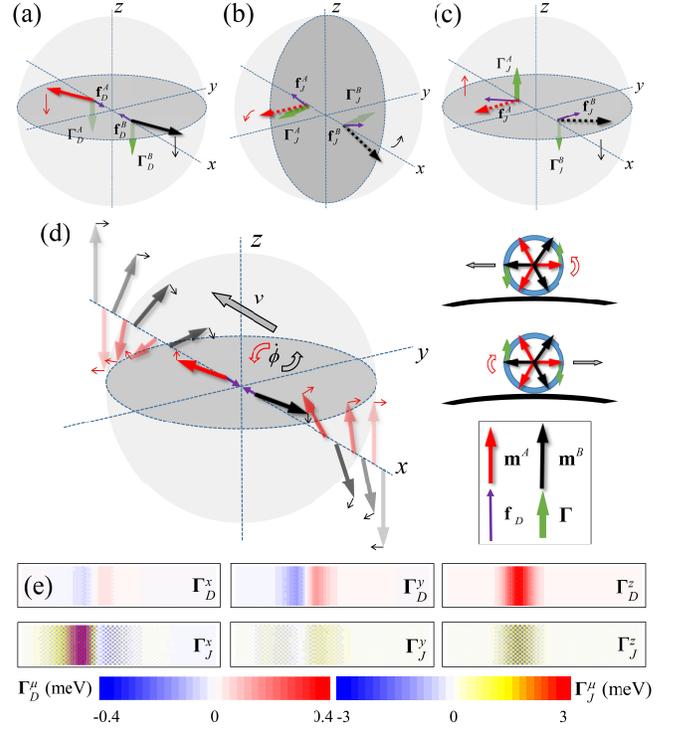}
\caption{Dynamic analysis around the domain wall. (a) After the polarization flipping, the DM effective fields ($\textbf{f}_D^{A/B}\sim\nabla\textbf{m}^{B/A}\times\textbf{D}$) are along the $x$-axis and antiparallel between neighbors, since $\nabla\textbf{m}^{A/B}$ are along the $z$-axis and antiparallel between neighbors. The DM torques (${\bf \Gamma}_D^{A/B}\sim\textbf{f}_D^{A/B}\times\textbf{m}^{A/B}$) are parallel between neighbors, and along the $z$-axis. (b) Such parallel ${\bf \Gamma}_D^{A/B}$ induce the canting between $\textbf{m}^{A/B}$ and following effective fields ($\textbf{f}_J^{A/B}\sim2J\textbf{m}^{B/A}$) from exchange and thus additional antiparallel in-plane rotational torques (${\bf \Gamma}_J^{A/B}\sim2J\textbf{m}^{B/A}\times\textbf{m}^{A/B}$). (c) Despite the identical amplitude of ${\bf \Gamma}_J^{A/B}$, the different amplitudes of $\textbf{m}^{A/B}$ lead to different rotational angles. Such asynchronous procession further induces additional torques ${\bf \Gamma}_J^{A/B}$, antiparallel between neighbors and with components along the $z$-axis. (d) Such $xz$-plane rotational torques lead to the domain wall motion. Noting the direction of motion, locked by the azimuthal angular velocity ($\dot{\phi}$) of moments. The cartoon of wheels is for intuitional understanding. (d) The distributions of three components of DM torques and net exchange torques, taking from above LLG simulation at $23$ ps. Indeed, ${\bf \Gamma}_D$ (${\bf \Gamma}_J$) are nearly parallel (antiparallel) between nearest neighbors. A movie of dynamics and torques is included in SM (dynamic.gif).}
\label{Torque}
\end{figure}

Then it is crucial to understand above dynamics, especially for the translational motion of domain wall, i.e., the ``rolling-downhill" process. In fact, to generate this ``rolling" effect, there are two essential conditions required for torques. First, there must be torques in the $xz$-plane. Second, these torques in the $xz$-plane should be (almost) antiparallel on neighbor sites A and B. The (almost) antiparallel condition can be easily satisfied by the exchange $J$ as a principle of action-and-reaction. Then the only key condition is to generate the torques in the $xz$-plane. With a tiny bias in the beginning, e.g. small canting of $\textbf{m}^{A/B}$ along the $y$-axis, the DM interaction can trigger the procession and lead to torques in the $xz$-plane, as sketched in Fig.~\ref{Torque}(a-d). The real torques in above LLG simulation [Fig.~\ref{Torque}(e)], confirm this analysis.

The dynamics of staggered moments around domain wall can also be analytically formulated as:
\begin{equation}
\frac{\partial\textbf{n}}{\partial{t}}=-v\nabla{\textbf{n}}+\dot{\phi}\hat{z}\times\textbf{n},
\end{equation}
where the first term in the right side describes the translational motion of domain wall and $v$ is the effective velocity; the second term describes the rotation  of $\textbf{n}$ around the $z$-axis and $\dot{\phi}$ is the azimuthal angular velocity. Here we assume all moments rotate synchronously, i.e., with an (almost) identical azimuth angle $\phi$ which is a proper approximation according to above LLG simulation at zero temperature.

Then Eq.~\ref{LLG} can be rewritten as a torque equilibrium equation for the whole lattice \cite{SM}:
\begin{equation}
\int_{-\infty}^{\infty}[\frac{\partial\textbf{n}}{\partial{t}}+\gamma^\prime\textbf{n}\times\textbf{f}-\frac{\alpha}{\sigma}(\textbf{n}\times\frac{\partial\textbf{n}}{\partial{t}})]dx=0,
\label{movement}
\end{equation}
where $\sigma$=$(m^A-m^B)/(m^A+m^B)$ is the ferrimagnetic ratio and $\gamma^\prime$=$-2\gamma/(m^A-m^B)$. Due to the uniform precession, the energy items of exchange interaction and magnetic anisotropy are almost unchanged during the dynamics. Thus we can consider the DM interaction only to the effective field torque. From Eq.~\ref{movement}, the motion $d$ and $\phi$ can be obtained as (more details can be found in SM \cite{SM}):
\begin{eqnarray}
&&d=\frac{\Delta\sigma}{\alpha}\atan[\sinh(t/Q)],\\
&&|\sin\phi|=\sech(t/Q),
\end{eqnarray}
where $Q=2\Delta(\alpha^{-1}\sigma+\alpha\sigma^{-1})/(\pi\gamma^{\prime}D_y)$. Then the characteristic time for the flip-flop process is $2Q$, which leads to a final movement distance:
\begin{equation}
d_s=\frac{\pi\Delta\sigma}{\alpha}.
\label{ds}
\end{equation}

Interestingly, the final distance is independent of strength of the DM interaction, in consistent with the LLG simulation [Fig.~S2(a) in SM \cite{SM}], but strongly relies on the damping constant and ferrimagnetic ratio. It is an advantageous fact since the SOC (and thus DM interaction) is generally not large for multiferroics based on $3d$ transition metal. The analytic solutions are shown in Fig.~\ref{md}(d-e) as curves, which match the simulation data quite well.

\textit{Discussions.} Although above simulations and analytical derivations have unambiguously demonstrated considerable CME via domain wall motion, there are several practical issues to be considered.

\begin{figure}
\includegraphics[width=0.48\textwidth]{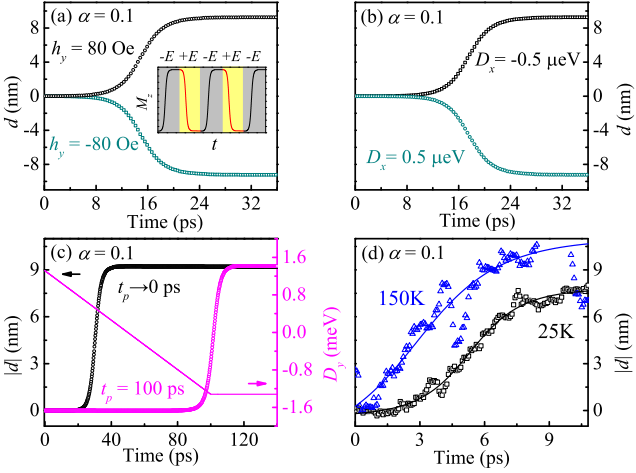}
\caption{Directional control of domain wall motion using (a) a transverse magnetic bias or (b) a longitudinal DM bias which can be driven by electric field along $x$-axis via the dielectric polarization. Insert of (a): reversible CME switched by AC electric field under fixed transverse magnetic field. (c) Left axis: motion of domain wall as a function of time when $D_y$ changes linearly: $D_y^0(1-2t/t_p)$ from $t=0$ to $t_p$. Various $t_p$'s are tested. Right axis: an example of $D_y$. (d) Simulations of domain wall motion at finite temperatures.}
\label{practice}
\end{figure}

First, the ``rolling-downhill" mechanism itself can not determine the motion direction, which depends on the bias of initial state. According to Eq.~\ref{movement}, the azimuthal angular velocity $\dot{\phi}$ and translational velocity $v$ are locked as:
\begin{equation}
v=-\frac{\Delta\sigma}{\alpha{C}}\dot{\phi}.
\label{relationship}
\end{equation}
Then a bias of magnetic field along the $y$-axis can control the initial rotational direction $\dot{\phi}$ and thus the motion direction, as proved in Fig.~\ref{practice}(a). In addition, a bias of electric field along the $x$-axis can also determine the motion direction, as proved in Fig.~\ref{practice}(b), which works via an additional DM component along the $x$-axis (i.e. $D_x$) from the dielectric polarization \cite{SM}. Thus, the magnetic domain wall motion can be driven by pure electric field without the help of magnetic field.

Second, for real materials, the reversal of $D_y$, i.e., the reversal of ferroelectric polarization, also needs time, which is beyond the sudden reversal approximation. Microscopically, the characteristic time of polarization switching would be in the range of $10-100$ ps \cite{li_ultrafast_2004, mankowsky_ultrafast_2017}. To account this point, a linear change of $D_y$ is considered with a given period $t_p$. As shown in Fig.~\ref{practice}(c), the motion of domain wall starts only after the sign reversal of $D_y$. It agrees with above results that the magnitude of $D_y$ is not important but its sign is essential. Therefore, the total time for the motion can be a simple sum of ionic relaxation plus spin dynamics. 

Third, all above simulations were done without any temperature effect. While in real practices, any device should work at finite temperatures. The temperature effect can be mimicked in the LLG simulations by adding a time-dependent random field to the effective field term and to solve the LLG equation by Heun method \cite{brown_thermal_1963,nowak_classical_2007,SM}. The amplitude of random field should be in proportional to the thermal energy. The LLG simulations of domain wall motion have been repeated at several temperatures. As shown in Fig.~\ref{practice}(d), the motions are similar to above zero-temperature one, with accelerated starts of the intermediate state due to the thermal fluctuation.

Fourth, although above work was done on a ferrimagnetic system, the conclusion can be straightforwardly extended to ferromagnetic and antiferromagnetic systems. According to Eq.~\ref{ds}, the motion distance $d_s$ is in proportional to $m_A-m_B$. Thus this motion does not exist in the antiferromagnetic cases. In contrast, the ferromagnetic cases should be similar to the ferrimagnetic ones by considering the sites A and B as a unit. Our LLG simulations confirm the existence and absence of domain wall motion in ferromagnetic and antiferromagnetic cases, respectively \cite{SM}. Although ferroelectric ferromagnets are rather rare, the ferroelectric ferromagnetic heterostructures are available, in which our mechanism also works, as long as the polarization can generate effective DM effect in proximate magnetic layers \cite{guo2017interface,guo_visualizing_2019}.

Last, our estimated dynamic characteristic time is in real unit, implying a fast process in real experiments. Of course, this characteristic time depends on the intensities of magnetic interactions of materials. And the coefficient $\alpha$ can be measured experimentally. For reference, for a ferromagnetic perovskite La$_{0.7}$Sr$_{0.3}$MnO$_3$, $\alpha$ was estimated as $0.01$ \cite{madon_room_2018}. For magnetic insulators (as required for ferroelectrics), $\alpha$ should be even lower since there is no dissipation from conductive electrons \cite{soumah_ultra-low_2018,kapelrud_spin_2013}. Then the effect of CME should be even stronger, since the motion distance of domain wall $d_s\propto1/\alpha$. For example, if $\alpha=0.001$ and other coefficients unchanged, the expected motion of ferrimagnetic domain wall can be close to $1$ $\mu$m, which can be visualized in magnetic force microscopes. For ferromagnetic domain wall, this CME effect can be even stronger [see Fig.~S2(c) in SM \cite{SM}], and thus easier to be visualized.

In conclusion, we proposed a physical mechanism based on spin dynamics to manipulate ferrimagnetic (or ferromagnetic) domain walls by switching the ferroelectric polarization, which can lead to significant converse magnetoelectric effect in nanoscale. Our mechanism relies on the chirality of domain wall, which is locked by the ferroelectric polarization via spin-orbit coupling. The reversal of chirality leads to significant translational motion of domain wall, resizing the magnetic domains. Our mechanism works well when spin-orbit coupling is small, and provides an alternative route to manipulate magnetism using voltage.

\begin{acknowledgments}
This work was supported by the National Natural Science Foundation of China (Grant No. 11834002). We thank the Tianhe-II of the National Supercomputer Center in Guangzhou (NSCC-GZ) and the Big Data Center of Southeast University for providing the facility support on the numerical calculations.
\end{acknowledgments}

\bibliographystyle{apsrev4-1}
\bibliography{myref}
\end{document}